\documentclass[12pt]{article}
\usepackage{epsfig}
\usepackage{graphicx}
\usepackage{color}
\usepackage{amsmath}

\usepackage{makeidx}
\makeindex

\renewcommand{\theequation}{\arabic{section}.\arabic{equation}}
\renewcommand{\theequation}{\arabic{equation}}

\stepcounter{footnote}

\makeatother
\def \baselinestretch{1.5}
\large \normalsize
\begin{document}
\setcounter{page}{1}
\setcounter{section}{0}

\normalsize

\def \footnotesep{1.75em}

\title{\normalsize \textbf {A PARETO SCALE-INFLATED OUTLIER MODEL AND ITS BAYESIAN ANALYSIS}} 

\author{\normalsize David PM Scollnik}

\date{}
\maketitle

\vspace*{-0.5in} 
\begin{center}
{\it Department of Mathematics and Statistics, 
University of Calgary\\ 
Calgary, Alberta, Canada\\ 
scollnik@ucalgary.ca} 
\end{center}
\vspace*{.2in}

\vspace*{-0.4in} 
\begin{center}
\normalsize{ABSTRACT}\\
\normalsize
\end{center}
This paper develops a Pareto scale-inflated outlier model. 
This model is intended for use when data from some standard Pareto distribution of interest is suspected to have been contaminated with 
a relatively small number of outliers from a Pareto distribution with the same shape parameter but with an inflated scale parameter. 
The Bayesian analysis of this Pareto scale-inflated outlier model is considered and its implementation using the Gibbs sampler is 
discussed. 
The paper contains three worked illustrative examples, two of which feature actual insurance claims data. 

\normalsize

\vspace*{+0.1in} 
\begin{center}
KEYWORDS
\end{center}
\begin{center}
Pareto distribution, 
Mixture model, 
Outliers, 
Insurance, 
Bayesian analysis, 
Gibbs sampler
\end{center}

\pagestyle{plain}
\renewcommand{\theequation}{\arabic{section}.\arabic{equation}}


\section{Introduction}
\label{intro}
The Pareto distribution is arguably one of the most popular and widely used of those in the class of continuous univariate distributions. 
Excellent overviews of the Pareto distribution are available in Arnold (1983) and Johnson, Kotz and Balakrishnan (1994). 
Vilfredo Pareto originally formulated it to describe the allocation of wealth among individuals, a situation in which a larger portion of 
the wealth in a society is owned by a smaller percentage of people therein. 
The Pareto distribution has since been used to profitably model many other situations, particularly those in which an equilibrium 
is found in the distribution of the ``small" values to the ``large". 
Some of the applications of the Pareto (and its related) distributions include modelling 
distributions of city population sizes, 
the occurrence of natural resources (e.g. size of oil reserves in oil fields), stock price fluctuations, 
size of firms, and error clustering in communication circuits 
(see Johnson \textit{et al.} (1994)). 
The Pareto distribution is also commonly used to model the severity of large casualty losses for certain lines of business 
such as fire and general liability, motor insurance, and workers compensation (e.g., McNeil (1997), Scollnik (2007), Schmutz 
and Doerr (1998)). 

This paper will develop a model based on the Pareto distribution for use in certain situations when it is feared that the data is 
contaminated with one or more outliers. 
An outlier may be thought of as an outlying observation that is numerically distant from the rest of the data. 
Or, harkening back to Grubs (1969), ``one that appears to deviate markedly from other members of the sample in which it occurs". 
A thick tailed Pareto distribution will as a matter of course generate occasional observations distant from the rest of the data. 
However, an outlier can also arise as an observation that does not come from the assumed default model. 
Depending upon the statistical approach taken, the distribution of the outliers generated by something other than the assumed 
default model may or may not be specified. 
For instance, as noted by an anonymous reviewer of a related paper of ours, in the field of robust statistics it is usually assumed 
that the main part of the data follows a model and the distribution of outliers is not specified. 
The aim there is to reduce the influence of the outliers on the estimation of the model for the main part of data, but not to model 
the outliers themselves.

In this paper, the approach taken is that most observations are from an assumed default Pareto($\alpha, \theta$) model with a 
certain threshold of $\theta$, but that occasional outliers are generated from a different Pareto model with a higher threshold, say, 
$\beta \theta$ with $\beta > 1$. 
We will refer to the parameter $\beta$ as a scale inflation factor. 
Note, if the expected value of the default Pareto distribution exists (i.e., if $\alpha > 1$) then the expected value of the alternative 
Pareto distribution is $\beta$ times as great. 
We will refer to the model developed in this paper as a Pareto scale-inflated outlier model. 
However, it could also be described as a two-component mixture of Pareto distributions model. 
Our aim in this paper is primarily to use this model to reduce the influence of a relatively small number of outliers as discussed 
above on the estimation of the model parameters, especially $\alpha$, for the main part of the data. 
This model certainly may also be utilized when a set of data is more evenly split between the default and the higher-threshold 
Pareto models. 
However, that scenario is not the main topic of this paper. 

Dixit and Jabbari Nooghabi (2011a) previously considered a Pareto based model with the presence of outliers and claimed that 
their ``work is the first in estimation in the Pareto distribution with outliers". 
In their paper, they let the set of $n$ random variables $( X_1, X_2, \ldots, X_n )$ represent claim amounts of a motor insurance 
company, and assume that $k$ of these ($k \geq 1$) claims are associated with some particular sort of vehicles (e.g. more expensive 
and / or more severely damaged) such that these claims are $\beta$ times higher than those of the standard (or typical) vehicles. 
Their assumption is that the claim amounts of the standard (or typical) vehicles are distributed with Pareto($\alpha, \theta$) probability 
density function (pdf) 
\begin{equation}
\label{model1}
f( x; \alpha, \theta ) = \frac{\alpha \, \theta ^ {\,\alpha} }{x ^ {\, \alpha + 1 }}, ~~~ 0 < \theta \leq x, ~ \alpha > 0, 
\end{equation}
and that those of the remainder (the outliers) have the Pareto($\alpha, \beta \theta$) pdf 
\begin{equation}
\label{model2}
f( x; \alpha, \beta, \theta ) = \frac{\alpha \, ( \beta \, \theta) ^ {\,\alpha} }{x ^ {\, \alpha + 1 }}, ~~~ 
0 < \beta \, \theta \leq x, ~ \alpha > 0, ~ \beta > 1, ~\theta > 0. 
\end{equation}
Dixit and Jabbari Nooghabi (2011a) assume that $\beta$, $\theta$, and $k$ (the number of outliers) are all known, and that 
$\alpha$ is unknown. 
Dixit and Jabbari Nooghabi (2011a) employ maximum likelihood estimation and uniformly minimum variance unbiased estimation. 
See Scollnik (2012) for a reexamination and correction of some of their reported results.  
Dixit and Jabbari Nooghabi (2011b) develop an extension of their model in which $\theta$ and $\beta$ may also be unknown. 
However, the estimation method they employ in this case (a combination of method of moments and least squares) can yield 
parameter estimates that are inconsistent with the observed data. 
See Scollnik (2012) for the details on this, as well as the derivation of a maximum likelihood estimation procedure. 
It is important to note that the random variables $( X_1, X_2, \ldots, X_n )$ are not independent in Dixit and Jabbari Nooghabi's 
models. For more on the nature of the dependence, see Dixit and Jabbari Nooghabi (2011a, page 342) and (2011b, page 819). 

The Pareto scale-inflated outlier model to be developed and discussed in this paper is more along the lines of the contaminated 
outlier models explored in Verdinelli and Wasserman (1991). 
In particular, 
and unlike the situation in Dixit and Jabbari Nooghabi's models, 
the observations will be assumed to be independent of one another given the model parameters. 
And, also as in Verdinelli and Wasserman (1991), this paper will develop a Bayesian statistical analysis using the Gibbs sampler. 
The estimation methodology used in this paper will always be the same, regardless of which particular model parameters are known 
and which are unknown (i.e. unlike the situations described above concerning estimation of the parameters in the Dixit and Jabbari 
Nooghab models). 
Neither the exact number of outliers, nor the probability that an observation is an outlier, will need to be known. 
The Bayesian approach will, however, allow prior information with respect to the number of outliers, or any model parameter, to be included 
in the analysis. 
The Bayesian approach will also allow the posterior inferences to be averaged over, or marginalized with respect to, the possible values 
of $k$. Predictive inferences incorporating parameter uncertainty are also available using this methodology. 

The Pareto scale-inflated outlier model will be defined and discussed in Section 2. 
Its Bayesian analysis using the Gibbs sampler will be developed in Section 3. 
This will be followed by three worked illustrative examples. 
The first example makes use of simulated data and appears in Section 4. 
The remaining two examples feature actual insurance claims data. 
Specifically, Section 5 considers a motor insurance claims data set and Section 6 addresses a medical insurance claims example. 

\section{The Pareto outlier model}
\label{jdl}
An early and well-studied form of outlier model is the contaminated location-shift normal. 
See, for example, 
Guttman, Dutter, and Freeman (1978). 
This is also one of the outlier models discussed in Verdinelli and Wasserman (1991). 
It assumes that the random variables $( X_1, X_2, \ldots, X_n )$ are a sample from the distribution with pdf of the form 
\begin{equation} 
f( x_i \,|\, \mu, \sigma^2, \epsilon, A_{i} ) = 
( 1 - \epsilon ) \, \phi( x_i \,|\, \mu, \sigma^2 ) + \epsilon \, \phi( x_i \,|\, \mu + A_i, \sigma^2 ), 
\end{equation}
where $\phi( x \,|\, \mu, \sigma^2 )$ is the normal density with mean $\mu$ and variance $\sigma^2$, and $\epsilon$ is the probability that the 
observation is from the normal distribution with its location shifted by an amount given by $A_i$. 
Given the model parameters, the $X_i$ are all independent of one another. 
Verdinelli and Wasserman (1991) re-express this model by introducing independent Bernoulli trials $\delta_i$, $i = 1, \ldots, n$, each with 
success probability $\epsilon$. Then, 
\begin{equation} 
f( x_i \,|\, \mu, \sigma^2, \epsilon, A_{i}, \delta_i ) = \phi( x_i \,|\, \mu + \delta_i A_i, \sigma^2 ). 
\end{equation}
Verdinelli and Wasserman (1991) assume standard conjugate priors for $\mu$ and $\sigma^2$ and assume that the $A_i$\,s are independent 
with identical zero mean normal prior distributions. They implement the corresponding Bayesian analysis using the Gibbs sampler. 

The Pareto scale-inflated outlier model considered in this paper is similar in spirit to the one above, but with some significant differences. 
Specifically, assume that the random variables are from the distribution with pdf 
\begin{equation} 
\label{model3} 
f( x_i \,|\, \alpha, \theta, \beta, \epsilon ) = 
( 1 - \epsilon ) \, \frac{\alpha \theta^{\alpha}}{x_i^{\alpha + 1}} \, \mathbf{I}( x_i - \theta ) + 
\epsilon \, \frac{\alpha ( \beta \theta )^{\alpha}}{x_i^{\alpha + 1}} \, \mathbf{I}( x_i - \beta \theta )
\end{equation}
in which $\alpha > 0$, $\theta > 0$, and $\beta > 1$. Here, $\mathbf{I}$ is the indicator function defined as 
\begin{equation}
\label{indicator1}
\mathbf{I}( y ) = 
\begin{cases}
1 ~~~ y \text{ $\geq$ 0},\\
0 ~~~ \text{otherwise}. 
\end{cases}
\end{equation}
The indicator functions arise in the definition of the model as the constituent Pareto distributions have different support. 
In this model, the outlying observations are seen to be coming from a scale-inflated Pareto distribution. 
The model can be re-expressed as 
\begin{equation} 
\label{model4}
f( x_i \,|\, \alpha, \theta, \beta, \epsilon, \delta_i ) = 
\frac{\alpha ( \beta^{\,\delta_i} \theta )^{\alpha}}{x_i^{\alpha + 1}} \, \mathbf{I}( x_i - \beta^{\,\delta_i} \theta ). 
\end{equation}
As before, the $\delta_i$ are independent Bernoulli random variables with an identical probability of success given by $\epsilon$. 
Note that the $X_i$ are conditionally independent of one another, and also of $\epsilon$, given the other model parameters. 

The Pareto scale-inflated outlier model differs from the contaminated location-shift normal model in a couple of ways, beyond 
the obvious that the former is built up using Pareto distributions whereas the latter uses normal. 
Another difference is that the scale-inflation parameter $\beta$ is assumed to be common for all observations, whereas the location 
shift parameter $A_i$ varies from observation to observation. 
This is simply due to the nature of the model we are constructing. 
That is, a common $\beta$ seems appropriate for many insurance contexts and in particular is an assumption that is appropriate for 
the illustrative examples to follow later in this paper. 
However, it would not be difficult to adjust the model to allow differing values of $\beta$, say $\beta_i$, for different observations. 
The analysis could still go forward using the methodology described in this paper with just a few changes. 
(For some insurance examples we have considered, this model adjustment did not greatly affect the overall analysis.) 
Another difference has to do with the support of the $X_i$. 
Under the contaminated location-shifted normal model, the support of these variables is independent of $\delta_i$, $\mu$, and $A_i$. 
But under the Pareto scale-inflated outlier model, the support of $X_i$ varies with the values of $\delta_i$, $\theta$, and $\beta$. 
This introduces subtleties and complications into the Pareto scale-inflated outlier model that do not exist in the contaminated location-shifted 
normal model. 
In particular, given the observed values of the $X_i$ variables, the varying support implies range restrictions on some model parameters. 
These restrictions must be monitored and incorporated in the implementation of the Gibbs sampler. 


In order to perform a Bayesian analysis of the Pareto scale-inflated outlier model defined above using the Gibbs sampler, we need to 
consider the selection of prior distributions for the model parameters and establish the form of the resulting full conditional posterior 
distributions. This is all discussed and illustrated in the following sections. 

We assume that readers are familiar with the basic ideas underlying the Gibbs sampler and other methods of Markov chain Monte 
Carlo (MCMC). 
An excellent review of these subjects, and of Bayesian inference in general, is available in Gelman et al. (2004). 
See Ntzoufras (2009) for another excellent overview. 
MCMC methods have attracted widespread use and attention in the statistics community and literature over the last two decades. 
Many papers featuring MCMC methods have also appeared in the actuarial literature in recent years. 
These include papers by 
Scollnik (2001),  
Verrall (2004), 
Ntzoufras et al. (2005), 
de Alba (2006), and 
Verrall (2007), 
to name just a few.

\section{Implementing the Bayesian analysis of the model}
\label{bay1}
Let $\boldsymbol{X} = ( X_1, X_2, \ldots, X_n )$ be a random sample from the Pareto scale-inflated outlier model \eqref{model3} and 
let $\boldsymbol{\delta} = ( \delta_1, \ldots, \delta_n )$. 
Then 

\begin{equation} 
\label{likd}
f( \boldsymbol{x} \,|\, \alpha, \theta, \beta, \epsilon, \boldsymbol{\delta} ) = 
\frac{\alpha^n ( \beta^{\,k} \theta^{\,n} )^{\alpha}}{\prod_{i=1}^n x_i^{\,\alpha + 1}} \, 
\prod_{i=1}^n \mathbf{I}( x_i - \beta^{\,\delta_i} \theta ) 
\end{equation}
with $\alpha > 0$, $\theta > 0$, $\beta > 1$, and where $k = \sum_{i = 1}^n \delta_i$. 

Assume that the model parameters, with the exception of $\epsilon$ and $\boldsymbol{\delta}$, are conditionally independent of one 
another \textit{a priori}. 
In this case, the posterior distribution for all of the model parameters is given by 
\begin{equation} 
\label{likd2}
f( \alpha, \theta, \beta, \epsilon, \boldsymbol{\delta} \, |\, \boldsymbol{x} ) \propto 
f( \alpha ) f( \theta ) f( \beta ) f( \epsilon ) f( \boldsymbol{\delta} \,|\, \epsilon ) \, 
\left( \frac{ \theta ^ {\,n} \beta ^ {\,k}}{\prod_{i = 1}^n x_i} \right) ^ \alpha \, 
\prod_{i=1}^n \mathbf{I}( x_i - \beta^{\,\delta_i} \theta ). 
\end{equation}
%
Recall, the $\delta_i$, $i = 1, \ldots, n$, are independent Bernoulli trials with common success probability $\epsilon$, \textit{a priori}, so that 
\begin{equation} 
\label{deltacond}
f( \boldsymbol{\delta} \,|\, \epsilon ) \propto 
\prod_{i=1}^n \epsilon^{\,\delta_i} (1 - \epsilon )^{1 - \delta_i} \,. 
\end{equation}

In order to implement a Gibbs sampler, we must first identify the form of the full conditional posterior distributions for each of the unknown 
model parameters. Next, we sample iteratively in an alternating fashion from each of the relevant full conditional posterior distributions in turn 
in order to obtain a random sample from the joint posterior. 
Details of this methodology can be found in any of many standard references now available, such as Gelman et al. (2004). 
The forms of the full conditional posterior distributions for the unknown model parameters are now established below, assuming some flexible 
but standard prior density specifications. 

The conjugate prior for $\alpha$ is the $\text{gamma}( a_1, a_2 )$ distribution (with mean $a_1 / a_2$ and variance $a_1 / a_2^2$). 
This choice of prior leads to a full conditional posterior distribution for $\alpha$ with a density given by 
\begin{align}
f( \alpha \,|\, \boldsymbol{x}, \theta, \beta, \boldsymbol{\delta} ) 
\nonumber
& \propto \alpha ^ {a_1 + n - 1} e ^ {- \alpha \, a_2} \left( \frac{ \theta ^ {\,n} \beta ^ {\,k}}{\prod_{i = 1}^n x_i} \right) ^ \alpha \\ 
& \nonumber \\ 
\label{alphacond1}
& \propto \alpha ^ {a_1 + n - 1} e ^ {- \alpha [ \, a_2 + \sum_{i = 1}^n \text{ln}( x_i ) - n \, \text{ln}( \theta ) - k \, \text{ln}( \beta ) ] }. 
\end{align}
This is readily identified as a $\text{gamma}( a_1 + n, a_2 + \sum_{i = 1}^n \text{ln}( x_i ) - n \, \text{ln}( \theta ) - k \, \text{ln}( \beta ) )$ distribution. 
Observe that this full conditional posterior distribution is independent of $\epsilon$, and depends on $\boldsymbol{\delta}$ only through the current 
value of $k$, i.e. in the current iteration of the Gibbs sampling algorithm. 
If we are confident that the mean (variance) of the model \eqref{model1} exists, then the corresponding restriction $\alpha >1$ ($\alpha > 2$) 
may be imposed on the prior distribution. Any such restriction will pass through and apply to the posterior as well. 

The form of the full conditional posterior distribution for $\epsilon$ is given by 
\begin{align}
\label{epsiloncond1}
f( \epsilon \,|\, \boldsymbol{x}, \alpha, \theta, \beta, \boldsymbol{\delta} ) 
\propto f( \epsilon ) \prod_{i=1}^n \epsilon^{\,\delta_i} (1 - \epsilon )^{1 - \delta_i} \,. 
\end{align}
This full conditional posterior distribution also depends on $\boldsymbol{\delta}$ only through the value of $k$. 
If the prior for $\epsilon$ is taken to be $\text{beta}( b_1, b_2 )$, then it is clear that the full conditional posterior distribution for $\epsilon$ is 
$\text{beta}( b_1 + k, b_2 + n - k )$. 

The full conditional posterior for each $\delta_i$, $i = 1, \ldots, n$, will be a discrete probability distribution. 
From \eqref{model4} 
and \eqref{deltacond}, its form is seen to be given by 
\begin{equation} 
\label{deltacond0}
f( \delta_i \,|\, \boldsymbol{x}, \alpha, \theta, \beta, \epsilon) 
\propto 
\epsilon^{\,\delta_i} (1 - \epsilon )^{1 - \delta_i} \beta^{\,\delta_i} \, \mathbf{I}( x_i - \beta^{\,\delta_i} \theta ) \,, 
\end{equation} 
where $\delta_i$ is equal to either 0 or 1. 
Observe that the $\delta_i$ are conditionally independent of one another, given the observed data and the other model parameters. 
From \eqref{deltacond0}, it follows that 
\begin{align} 
\label{deltacond1}
Pr( \delta_i = 0 \,|\, \boldsymbol{x}, \alpha, \theta, \beta, \epsilon) & = 
\begin{cases}
1 ~~~ & \text{if $\theta \leq x_i < \beta \theta$}\\
\frac{1 - \epsilon}{1 - \epsilon \,+\, \beta^\alpha \epsilon} ~~~ & \text{if $\beta \, \theta \leq x_i$}
\end{cases}
\end{align} 
and 
\begin{align} 
\label{deltacond2}
Pr( \delta_i = 1 \,|\, \boldsymbol{x}, \alpha, \theta, \beta, \epsilon ) & = 
\begin{cases}
0 ~~~ & \text{if $\theta \leq x_i < \beta \theta$}\\
\frac{\beta ^ \alpha \epsilon}{1 - \epsilon \,+\, \beta^\alpha \epsilon} ~~~ & \text{if $\beta \, \theta \leq x_i$}, 
\end{cases}
\end{align} 
for $i = 1, 2, \ldots, n$. 
Note that the conditional posterior probabilities for $\delta_i$ do not depend upon the precise value of $x_i$, only upon whether or not 
$x_i < \beta \, \theta$ or $x_i \geq \beta \, \theta$. 
As the Gibbs sampler proceeds, the value of $k = \sum_{i = 1}^n \delta_i$ can also be monitored at the end of each iteration in order to develop 
posterior inferences with respect to the number of outliers. 

If the values of the parameters $\theta$ and $\beta$ are fixed and known, 
e.g., as in the example contained in Dixit and Jabbari Nooghabi (2011a), 
then the Gibbs sampler can be implemented using only 
\eqref{alphacond1}, 
\eqref{epsiloncond1}, 
\eqref{deltacond1} and \eqref{deltacond2}. 
When the values of the parameters $\theta$ and $\beta$ are fixed and known, it also follows from \eqref{deltacond1} and \eqref{deltacond2} 
that any observation $x_i$ below $\beta \, \theta$ has zero marginal posterior probability of being an outlier, whereas any observation greater 
than or equal to $\beta \, \theta$ has the same marginal posterior probability of being an outlier as any other such observation. 
This is simply a consequence of the assumed Pareto outlier model when $\theta$ and $\beta$ are fixed and known. 
This result may or may not be appropriate in a particular application. 

Recall that the motivation in Dixit and Jabbari Nooghabi (2011a) was to develop a model assuming that claims associated with some particular 
sort of special vehicles are exactly $\beta$ times higher than those of standard (or typical) vehicles, with $\beta$ a known value ($\theta$ was also 
assumed to be known). 
However, $\beta$ is not likely to be known precisely in practice. 
Furthermore, it may be useful to rank the observations (vehicle claims) according to how probable they are, \textit{a posteriori}, of being outliers 
(with the understanding that a larger observation should have a correspondingly larger such posterior probability). 
For instance, vehicles associated with posterior probabilities above some set level may be targeted for inspection, either to identify possible fraud 
or to determine whether this particular type of vehicle is being properly classified. 
Allowing $\beta$ to vary addresses both of these issues. 

Let $f( \beta )$ denote the prior distribution of $\beta$, and assume that $\beta > \beta^* \geq 1$ where $\beta^*$ is some assumed known 
lower limiting value. 
Then the full conditional posterior distribution of $\beta$, from \eqref{likd2}, is of form 
\begin{align}
\label{betacond1}
f( \beta \,|\, \boldsymbol{x}, \alpha, \theta, \boldsymbol{\delta}, \epsilon ) 
& \propto f( \beta ) \, \beta ^ {\, \alpha k} \,
\prod_{i=1}^n \mathbf{I}( x_i - \beta^{\,\delta_i} \theta ) \,. 
\end{align}
This may be written as 
\begin{align}
\label{betacond2}
f( \beta \,|\, \boldsymbol{x}, \alpha, \theta, \boldsymbol{\delta}, \epsilon ) 
& \propto f( \beta ) \, \beta ^ {\, \alpha k} 
\end{align}
where $1 \leq \beta ^ * < \beta$ when $k = \sum_{i = 1}^n \delta_i = 0$, 
and $1 \leq \beta ^ * < \beta \leq x ^ * / \theta$ where 
$x ^ * = \min\limits_{i \ni \delta_i = 1}( x_i )$ 
%
%
%
(i.e. the smallest $x_i$ for which $\delta_i = 1$) when $k = \sum_{i = 1}^n \delta_i \geq 1$. 
Of course, the value of $x ^ *$ may and typically will vary from iteration to iteration of the Gibbs sampler. 
For the examples later in this paper, we will assign $\beta$ a shifted exponential prior distribution such that 
\begin{align}
\label{betaprior1}
p( \beta ) 
& = \lambda \, e^{- \lambda ( \beta - \beta^* )}, 
\end{align}
with $\beta > \beta^* \geq 1$ and $\lambda = 1$. 
This leads to a shifted and sometimes (i.e., when $k \geq 1$) truncated from above exponential full conditional posterior distribution for 
$\beta$, with the truncation as previously described. 

Finally, let $f( \theta )$ denote the prior distribution of $\theta$ as before. 
Then the full conditional posterior distribution of $\theta$, from \eqref{likd2}, is of form 
\begin{align}
\label{thetacond1}
f( \theta \,|\, \boldsymbol{x}, \alpha, \beta, \boldsymbol{\delta}, \epsilon ) 
& \propto f( \theta ) \, \theta ^ {\, \alpha n} \,
\prod_{i=1}^n \mathbf{I}( x_i - \beta^{\,\delta_i} \theta ) \,. 
\end{align}
This may be written as 
\begin{align}
\label{thetacond2}
f( \theta \,|\, \boldsymbol{x}, \alpha, \beta, \boldsymbol{\delta}, \epsilon ) 
& \propto f( \theta ) \, \theta ^ {\, \alpha n} 
\end{align}
where $0 < \beta^{\,\delta_i} \theta \leq x_i$ for all $i$ or, more concisely, 
$0 < \theta < \min\limits_{i}( x_i / \beta^{\,\delta_i})$. 
For the examples later in this paper, we will assign $\theta$ a $\text{gamma}( t_1, t_2 )$ prior distribution. 
This leads to a truncated $\text{gamma}( t_1 + \alpha \, n, t_2 )$ full conditional posterior distribution for $\theta$, with the truncation as 
previously described.

\section{An example with simulated data} 
\label{synth}
In order to illustrate the Pareto scale-inflated outlier model \eqref{model3} and its Bayesian analysis, we first explore its application in the context 
of a simulated data set. 
Sixteen observations were simulated from the Pareto($\alpha, \theta$) model \eqref{model1} using $\alpha = 2.5$ and 
$\theta = 50000$. These represent standard (or typical) claim amounts. 
Four observations were simulated from the scale-inflated Pareto($\alpha, \beta \theta$) model \eqref{model2} using $\alpha$ and $\theta$ 
as above, and with $\beta = 3$. These represent the outlier claim values. 
The complete set of observations are given below (the outliers are the final four entries in the last line): 

\vspace{0.25in} 
\indent
\ \ \ \ \ \ \ \ \ 57,726, \ \ \ \ \ 51,806, \ \ \ \ \ 82,475, \ \ \ \ \ 75,840, \ \ \ \ \ 86,115 \\
\indent
\ \ \ \ \ \ \ \ \ 140,691, \ \ \ 53,960, \ \ \ \ \ 57,176, \ \ \ \ \ 66,577, \ \ \ \ \ 81,512, \\
\indent
\ \ \ \ \ \ \ \ \ 57,099, \ \ \ \ \ 71,053, \ \ \ \ \ 56,012, \ \ \ \ \ 50,291, \ \ \ \ \ 59,197, \\
\indent
\ \ \ \ \ \ \ \ \ 51,918, \ \ \ \ \ 170,781, \ \ \ 161,296, \ \ \ \ 330,773, \ \ \ 219,582. \\

\noindent
This example will proceed under the assumption that the true value of $\theta$ is known. 
Interest is primarily in the estimation of the parameter $\alpha$ as it is the single parameter remaining that determines the distribution 
\eqref{model1} for the standard claims.   

For this illustrative Bayesian analysis we adopt the following prior specification. 
The parameter $\alpha$ is assigned a relatively diffuse, or noninformative, $\text{gamma}( 0.001, 0.001 )$ prior so that it has a mean of one 
and a very large variance. 
The parameter $\beta$ is assigned a shifted exponential prior distribution as in \eqref{betaprior1}. 
Finally, the parameter $\epsilon$ is assigned a beta$( b_1, b_2 )$ prior distribution with $b_1 = 0.1842$ and $b_2 = 3.5$. 
These last two values come from Verdinelli and Wasserman (1991) and their review of some of the literature pertaining to outlier models. 
This specification assigns $\epsilon$ a prior mean of 0.05, and assigns any observation ``less than half a chance of being an outlier with high 
probability" \textit{a priori} (Verdinelli and Wasserman, 1991, page 109). 
Specifically, this high prior probability is $\mathrm{Pr}( \epsilon < 0.5 ) = 0.99$. 
A different, and arguably more informative, prior density specification will be considered in the next example. 

\begin{figure}[!th]
\centering
\includegraphics[width=1.0\textwidth]{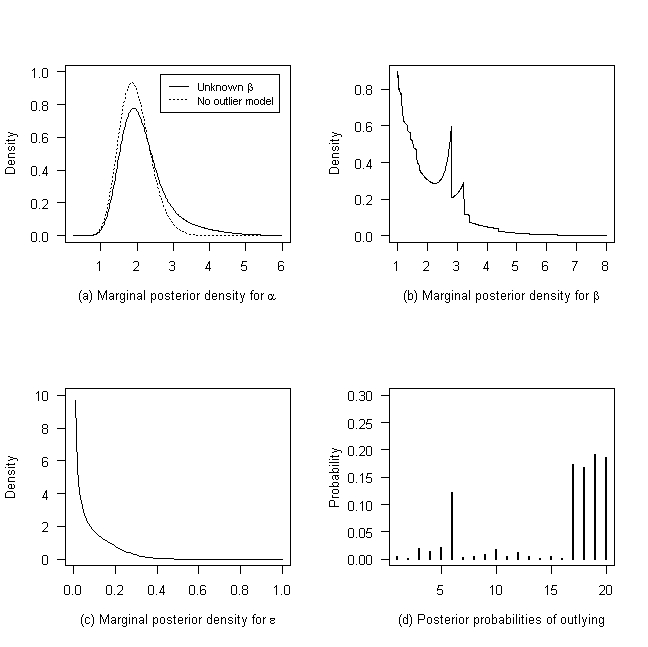}
\caption{Marginal posteriors for $\alpha$, $\beta$, $\epsilon$, and $\boldsymbol{\delta}$.}
\label{synth1a} 
\end{figure}

This Bayesian analysis was implemented using a Gibbs sampler constructed using the full conditional posterior distributions developed in 
Section 3. 
The summary inferences discussed below are based on a total of 200,000 kept iterations (following burn-ins of 10,000 iterations) of this Gibbs 
sampler. 
The parameter $\alpha$ has posterior mean of 2.193 and posterior standard deviation of 0.687. 
The posterior density of $\alpha$ is plotted with a solid line in Figure \ref{synth1a}(a). 
For comparison, the posterior density of $\alpha$ that results when the basic Pareto model \eqref{model1} (i.e. a Pareto model with no 
special accommodation made for outliers) is applied to the data in conjunction with the same prior for $\alpha$ as before is also given. 
This posterior density for $\alpha$ is of the same form as that given by \eqref{alphacond1} when $\beta = 1$. 
That is, the resulting posterior for $\alpha$ in the case of no outliers is 
$\text{gamma}( 20.001, 10.23953 )$. 
This density is plotted in Figure \ref{synth1a}(a) with a dotted line. 
From a comparison of the two posterior density plots, it is clear that the Pareto scale-inflated outlier model shifts more of the posterior mass 
towards the true value of $\alpha$, i.e. 2.5. 
This leads, in turn, to more reasonable and accurate statements about the predictive distribution of the future standard claim amounts.

\footnotesize
\begin{table}[tbp]
\begin{center}
\centerline{Table 1.} 
\centerline{Quantiles of the predictive distribution for standard claims under various models.} 
\vspace*{.1in}
\begin{tabular}{ @{\extracolsep{\fill}} | c c c c c | @{\extracolsep{0pt}} } 
\hline 
\, Pareto model \, & Median & 75\% & 90\% & 95\% \\ 
\hline
Actual & 65,975 & 87,055 & 125,594 & \, 165,722 \, \\ 
Scale-shifted & 69,364  & 98,351 & 158,719 & 232,623 \\ 
Basic & 71,749 & 104,260 & 174,328 & 261,562 \\ 
\hline 
\end{tabular} 
\end{center}
\end{table} 
\normalsize

Table 1 lists a number of the quantile values for various predictive claim amount models. 
The first line of quantiles correspond to those of the actual Pareto($\alpha, \theta$) model \eqref{model1} with $\alpha = 2.5$ and 
$\theta = 50000$. Recall, this is the model that generated the sixteen simulated standard claims. 
The second line of quantiles correspond to those of the Bayesian predictive distribution for the standard claims under the Pareto scale-inflated 
outlier model \eqref{model3}. 
This predictive distribution is defined by \eqref{model1} averaged over the posterior distribution of $\alpha$ that results under the Pareto 
scale-inflated outlier model analysis. 
The third line of quantiles in Table 1 correspond to those of the Bayesian predictive distribution arising when the basic Pareto model 
\eqref{model1} (i.e. a Pareto model with no special accommodation made for outliers) is applied to the data. 
It is apparent from Table 1 that the Pareto scale-inflated outlier model did a better job in this example of estimating these quantiles than did 
the basic Pareto model (which makes no allowance for outlying observations from a scale-inflated Pareto model). 
That the quantile values for the predictive distribution associated with the Pareto scale-shifted outlier model analysis are still slightly greater 
than the corresponding quantiles associated with the actual model is to be expected. 
This is a reflection of the effect of the parameter uncertainty in the former versus the parameter certainty in the latter. 

\begin{figure}[!th]
\centering
\includegraphics[width=1.0\textwidth]{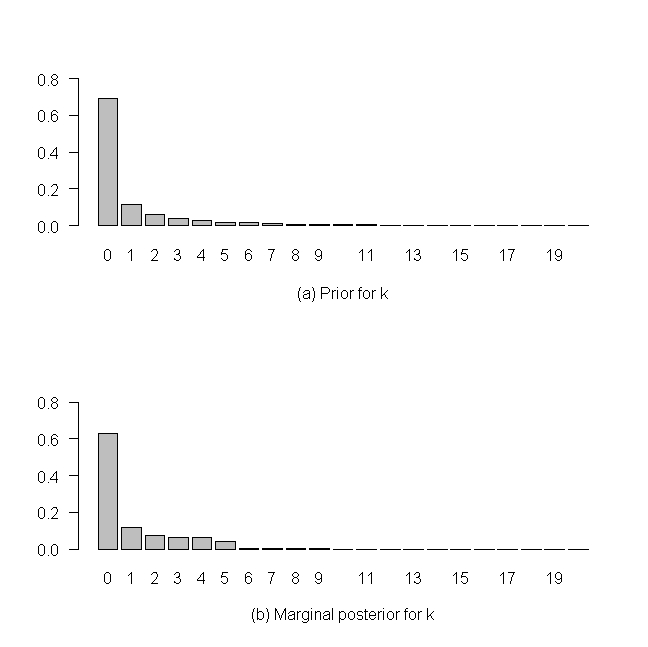}
\caption{Marginal prior and posterior probability mass functions for k.}
\label{synthk} 
\end{figure}

The parameter $\beta$ has posterior mean of 2.133 and posterior standard deviation of 0.954. 
The posterior density of $\beta$ is plotted in Figure \ref{synth1a}(b). 
Note that the bumps and dips in this marginal posterior density plot are a consequence of the varying range restrictions on $\beta$ that go 
along with (3.16), and correspond to the possible values of $x^* / \theta$. 
The posterior density for $\epsilon$ and the posterior probabilities than given observations are outliers are presented in the bottom left and 
bottom right graphs in Figure \ref{synth1a}, respectively. 
Observe that the four observations associated with the largest posterior probabilities of being outlying observations are, in fact, the known outliers. 
Figure \ref{synthk} contains a plot of the posterior probability function for $k$, i.e. the number of outliers. 
The mode of this posterior distribution is at $k = 0$, the same as the mode of the prior. 
However, the posterior clearly assigns more probability than the prior to the event that $k > 0$.

\section{An illustrative motor insurance example}
\label{motor}
The Pareto scale-inflated outlier model and its Bayesian estimation will now be considered in the context of a motor insurance example using the 
data from Dixit and Jabbari Nooghabi (2011a). 
This example involves an insurance company in Iran that provides motor insurance as one of its services. 
Claim amounts vary according to the damage to the vehicles, and the vehicles themselves are of different (and in some cases very high) costs. 
The company assumed that claims of the most expensive and severely damaged vehicles (i.e. the outliers) are 1.5 times higher than those 
of the standard (or typical) vehicles. 
A random sample of size 20 of the claim amounts (in Iranian Rials) from the year 2008 is available, and is given below: 

\vspace{0.25in} 
\indent
\ \ \ \ \ \ \ \ \ 750,000, \ \ \ \ \ 780,000, \ \ \ \ \ \ 630,000, \ \ \ \ \ 1,750,000, \ \ \ 1,450,000 \\
\indent
\ \ \ \ \ \ \ \ \ 3,000,000, \ \ \ 7,650,000, \ \ \ 4,210,000, \ \ \ 890,000, \ \ \ \ \ \ 950,000, \\
\indent
\ \ \ \ \ \ \ \ \ 1,240,000, \ \ \ 1,800,000, \ \ \ 1,630,000, \ \ \ 9,020,000, \ \ \ 4,750,000, \\
\indent
\ \ \ \ \ \ \ \ \ 3,250,000, \ \ \ 1,135,000, \ \ \ 1,326,000, \ \ \ 1,280,000, \ \ \ 760,000. \\
 

\noindent
Dixit and Jabbari Nooghabi (2011a) note that claims of at least 500,000 Rials can be made and that claims below 500,000 Rials are not entertained. 
So, in this example, $\beta$ = 1.5 and $\theta$ = 500,000. 
The parameters $\alpha$ and $\epsilon$ (as well as $\boldsymbol{\delta}$) are unknown, 
and the main objective in this example is to develop posterior inference concerning the parameter $\alpha$. 
This will be compared to the posterior inference concerning $\alpha$ that results in a case when $\beta$ is not precisely known, and also to the 
posterior inference resulting in the instance that a basic Pareto model with no outliers is applied to the claims. 
In order to implement any Bayesian analysis, we must first specify the priors for the unknown parameters. 

For many lines of property and casualty insurance, values of $\alpha$ are typically in the range from $\approx 0.8$ to $\approx 2.5$ (e.g., see 
Schmutz and Doerr, 1998). 
Of course, values of $\alpha$ outside of this range are also possible. 
As the claim amounts in this example relate to motor insurance, so that the possible claim amounts are relatively constrained and certain not to be 
incredibly catastrophic, it is quite reasonable to assume that $\alpha >1$ in order so that the mean of \eqref{model1} exists. 
Indeed, it is not uncommon for values of $\alpha$ in the case of motor insurance to exceed 2. E.g., see Rosenbaum (2011). 
With all of this in mind, for the purpose of this illustrative example the prior distribution for the parameter $\alpha$ is taken to be a $\text{gamma}( 10, 5 )$ 
truncated below at 1. 
The mean of this prior distribution is 2.038 and its standard deviation 0.608. 

\begin{figure}[!th]
\centering
\includegraphics[width=1.0\textwidth]{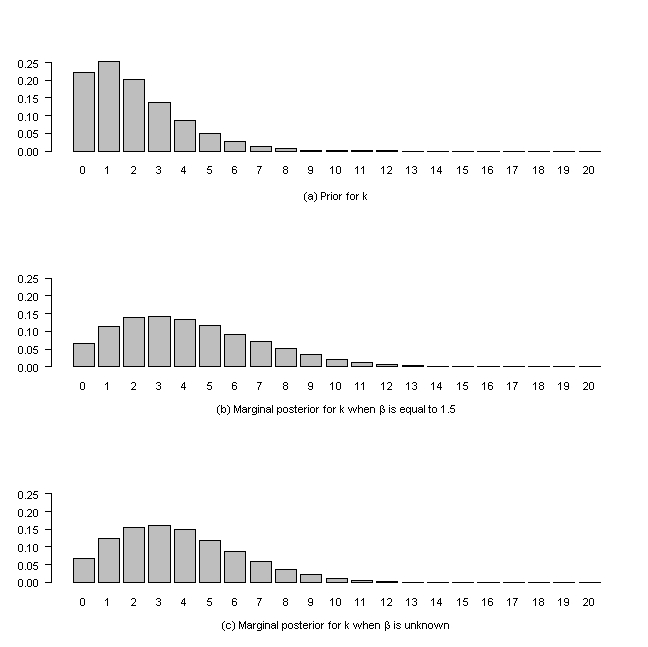}
\caption{Marginal prior and posterior probability mass functions for k.}
\label{motor1k}
\end{figure}

The discussion and context of the motor insurance data set in Dixit and Jabbari Nooghabi (2011a) suggests that the number of outliers (i.e., $k$) 
in this sample of size $n = 20$ should be relatively small (e.g., $k \approx 1 \, \text{to} \, 4$). 
Given $\epsilon$, the conditional prior distribution for the number of outliers in a sample of size 20 is binomial$( 20, \epsilon )$. 
Recall, the prior for $\epsilon$ is beta$( b_1, b_2 )$. This implies a beta-binomial marginal prior distribution on $k$. 
As this example involves an insurance company, it is reasonable to assume that actuaries or other knowledgeable experts at the company can 
make use of internal company and / or industry wide knowledge and / or insurance statistics in order to fashion informative \textit{a priori} 
statements about $k$. 
Assume their prior determination is that 2 outliers are expected in the sample (of size 20), and that 5 or less outliers should occur with 95\% probability. 
Given the previously mentioned beta-binomial distribution on $k$, the values $b_1 = 2.17484$ and $b_2 = 19.57356$ are consistent with this prior 
opinion. 
The corresponding prior mean of $\epsilon$ is $0.1$ and its prior standard deviation is approximately $0.0629$. 
The resulting marginal prior discrete distribution for $k$ 
is displayed in Figure \ref{motor1k}(a). 

\begin{figure}[!th]
\centering
\includegraphics[width=1.0\textwidth]{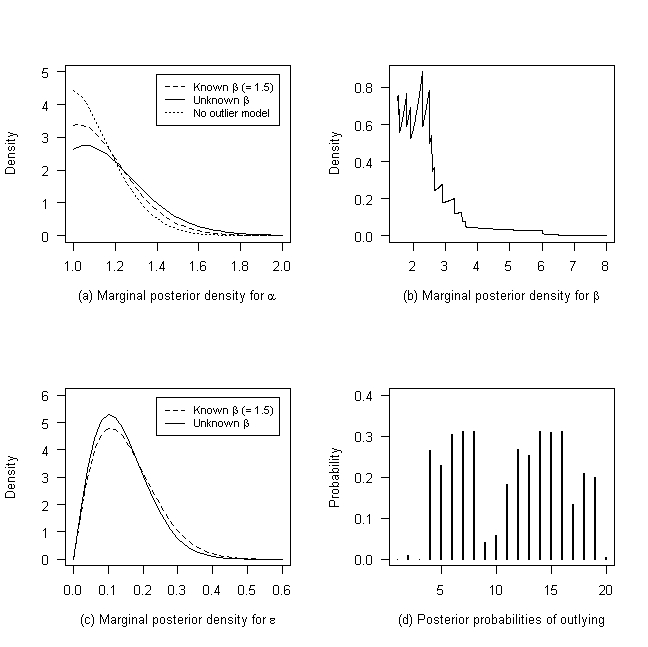}
\caption{Marginal posteriors for $\alpha$, $\beta$, $\epsilon$, and $\boldsymbol{\delta}$.}
\label{motor1} 
\end{figure}

The first Bayesian analysis was performed assuming that $\beta$ was fixed and set at 1.5. 
The parameter $\alpha$ has posterior mean of 1.188 and posterior standard deviation of 0.145. 
The parameter $\epsilon$ has posterior mean of 0.153 and posterior standard deviation of 0.086. 
The marginal posterior density of $\alpha$ and that of $\epsilon$ are both shown in Figure \ref{motor1} (the densities corresponding to the value of 
$\beta$ being known and equal to 1.5). 
From what was said in Section 3 concerning the conditional posterior probabilities of $\delta_i$, it follows that the marginal posterior probability of any 
given observation $x_i$ greater than $\beta \theta = 750,000$ being an outlier will be the same for all such observations. 
This follows as it is equal to the value of the corresponding outlier probability given in \eqref{deltacond2} marginalized over the simulated values of the 
other parameters. This estimated probability is approximately equal to 0.221. 
The posterior discrete distribution of $k$ (assuming $\beta = 1.5$) 
is illustrated in Figure \ref{motor1k}(b). 
The random variable $k$ has a posterior mean of 4.211, a posterior standard deviation of 2.795, and a posterior median equal to 4. 
Note that the simulation based inferences described above (and below) in this Section are all based on a total of 200,000 kept iterations (following 
burn-ins of 10,000 iterations) of a Gibbs sampler constructed as indicated in Section 3. 
Of course, each different Bayesian analysis was implemented using Gibbs sampler runs of its own. 

The second Bayesian analysis was performed assuming that the precise value of $\beta$ was unknown, but above some known limit. 
As in Section 3, assume that $\beta$ is assigned a shifted exponential distribution such that 
\begin{align}
\label{betaprior1b}
p( \beta ) 
& = \lambda \, e^{- \lambda ( \beta - \beta^* )}, ~~~ \beta \geq \beta^* > 1. 
\end{align}
We take $\beta^* = 1.5$ and $\lambda = 1$. 
This says that outliers have claims $\beta$ times higher than standard vehicles where $\beta$ is some unknown value, but one that is known to be 
at least $\beta^* = 1.5$. 
Under this second Bayesian analysis, 
$\alpha$ has a posterior mean of 1.228 and a posterior standard deviation of 0.176, 
$\beta$ has a posterior mean of 2.498 and a posterior standard deviation of 0.962, and 
$\epsilon$ has posterior mean of 0.141 and posterior standard deviation of 0.078. 
The marginal posterior densities for the parameters $\alpha$, $\beta$, and $\epsilon$ are plotted in Figure \ref{motor1}. 
The first of these plots shows that the marginal posterior density of $\alpha$ is concentrated more heavily on larger values of $\alpha$, compared to 
the first analysis in which $\beta$ was known. 
Note that the bumps and dips in the marginal posterior density plot for $\beta$ are a consequence of the varying range restrictions on $\beta$ that 
go along with \eqref{betacond1}, and correspond to the possible values of $x^* / \theta$. 
The marginal posterior probabilities that the different observations are outliers are also shown in Figure \ref{motor1}. 
These are obtained as the conditional outlier probabilities given in \eqref{deltacond2} marginalized over the simulated values of the other parameters. 
As $\beta$ is no longer assumed to be fixed, unlike in the previous analysis, the posterior outlier probabilities now vary from observation to observation. 
As was remarked earlier, an insurance company may be interested in flagging a vehicle with a posterior probability of being an outlier above some set 
high level for further examination, either to identify possible fraud or to determine whether this particular vehicle is being properly classified.
The posterior discrete distribution of $k$ when $\beta$ is unknown is illustrated in Figure \ref{motor1k}(c). 
The random variable $k$ has a posterior mean of 3.711, a posterior standard deviation of 2.391, and a posterior median equal to 4. 
It is apparent that the marginal posterior distribution for $k$ is less dispersed and concentrated more on the smaller values in this analysis, than 
it was in the previous one (when $\beta$ was assumed equal to 1.5). 

As previously remarked, for the sake of comparison a Bayesian analysis of the basic Pareto model (with no outliers), i.e. \eqref{model1}, 
applied to the data was also carried out. 
The posterior density for $\alpha$ under this analysis is plotted in Figure \ref{motor1}(a) with a dotted line. 
Comparing the three posterior density curves for $\alpha$, it is apparent that the Bayesian analyses associated with the Pareto scale-inflated outlier 
model places more posterior probability on larger values of $\alpha$ 
(especially so with the model when $\beta$ is unknown, with $\beta > 1.5$). 
This should lead, as in the previous example, to more sensible estimates of $\alpha$ and 
to more reasonable and accurate statements about the predictive distribution of future standard claim amounts and hence more reasonable motor 
insurance premiums for the standard vehicles. 
The effect on the predictive distribution is illustrated in Table 2, which lists a number of the quantile values for the predictive claim amount distributions 
associated with the standard vehicles that result under the three analyses described above.

\footnotesize
\begin{table}[tbp]
\begin{center}
\centerline{Table 2.} 
\centerline{Quantiles of the predictive distribution for standard motor claims under various models.} 
\vspace*{.1in}
\begin{tabular}{ @{\extracolsep{\fill}} | c c c c c | @{\extracolsep{0pt}} } 
\hline 
\, Pareto model \, & Median & 75\% & 90\% & 95\% \\ 
\hline
Scale-shifted $(\beta > 1.5)$ & 882,127 & 1,571,983 & 3,408,270 & \, 6,267,162 \, \\ 
Scale-shifted $(\beta = 1.5)$ & 902,218 & 1,632,503 & 3,598,453 & 6,546,247 \\ 
Basic & 912,938 & 1,680,812 & 3,757,930 & 6,909,201 \\ \hline 
\end{tabular} 
\end{center}
\end{table} 
\normalsize

\begin{figure}[!th]
\centering
\includegraphics[width=1.0\textwidth]{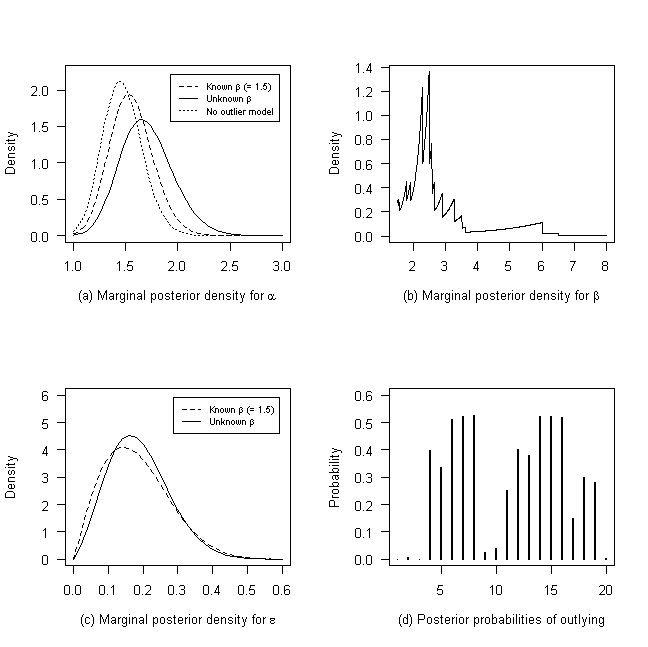}
\caption{Marginal posteriors for $\alpha$, $\beta$, $\epsilon$, and $\boldsymbol{\delta}$.}
\label{motor2} 
\end{figure}

\begin{figure}[!th]
\centering
\includegraphics[width=1.0\textwidth]{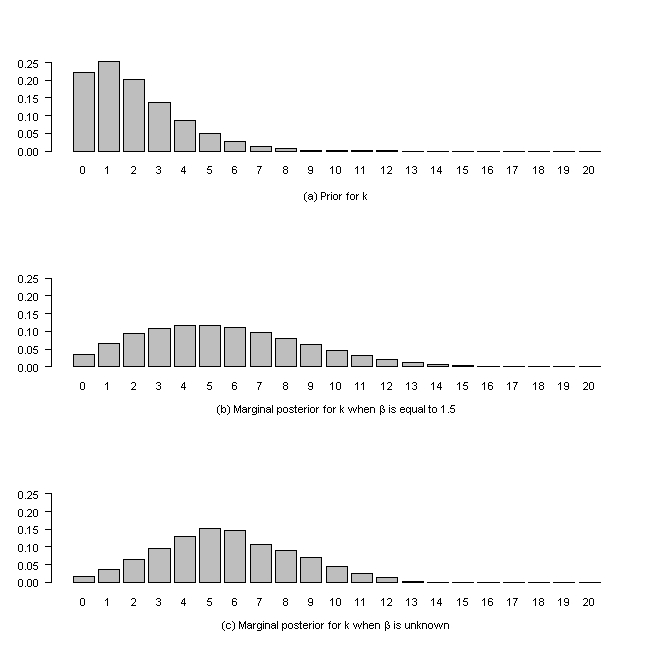}
\caption{Marginal prior and posterior probability mass functions for k.}
\label{motor2k}
\end{figure}

Recall, the Bayesian analyses of the basic Pareto and two Pareto scale-inflated outlier models above (i.e. when $\beta = 1.5$ and when $\beta > 1.5$) 
assumed that the prior distribution for $\alpha$ was $\text{gamma}( 10, 5 )$ truncated below at 1.
This was an informative prior but is perhaps not as informative as may often be available in practice, especially for a motor line of insurance. 
For illustrative purposes, we also considered the Bayesian analyses of these models when the prior for $\alpha$ was $\text{gamma}( 40, 16 )$ truncated 
below at 1.
This prior is significantly less dispersed than the earlier one and concentrates the prior probability of $\alpha$ fairly symmetrically around about 2.5. 
It also assigns  about 80\% of the prior probability to the interval between the values 2 and 3. 
The resulting posterior distributions for the various parameters under the different models appear in Figures \ref{motor2} and \ref{motor2k}. 
Note that the posterior marginals for $\alpha$ are shifted under the different models in the same way as in the previous set of analyses in this Section. 
Table 3 lists various quantile values for the predictive claim amount distributions associated with the standard vehicles that result under the different models, 
using this revised prior distribution on $\alpha$.



\footnotesize
\begin{table}[tbp]
\begin{center}
\centerline{Table 3.} 
\centerline{Quantiles of the predictive distribution for standard motor claims under various models.} 
\vspace*{.1in}
\begin{tabular}{ @{\extracolsep{\fill}} | c c c c c | @{\extracolsep{0pt}} } 
\hline 
\, Pareto model \, & Median & 75\% & 90\% & 95\% \\ 
\hline
Scale-shifted $(\beta > 1.5)$ & 755,085 & 1,142,615 & 1,996,918 & \, 3,070,095 \, \\ 
Scale-shifted $(\beta = 1.5)$ & 781,865 & 1,237,816 & 2,271,377 & 3,586,312 \\ 
Basic & 803,536 & 1,294,355 & 2,460,126 & 4,049,355 \\ 
\hline 
\end{tabular} 
\end{center}
\end{table} 
\normalsize

\section{An illustrative medical insurance example}
\label{medical}
The Pareto scale-inflated outlier model and its Bayesian estimation will now be considered in the context of a medical insurance example using  
data from Dixit and Jabbari Nooghabi (2011b). 
In this example, the values of $\alpha$, $\theta$, and $\beta$ are all unknown. 
This example involves an insurance company in Iran that provides medical insurance as one of its services. 
Claims may be made by passengers involved in a motor accident for medical expenses related to injuries sustained therein. 
Dixit and Jabbari Nooghabi (2011b) note that the amount of compensation (in Iranian Rials) is to be at least $\theta$ as claims less than this amount 
are not reasonable to claim. 
Claim amounts vary according to factors such as the type and nature of the injury. 
Most claims are near the value of $\theta$, which is assumed to be at (or near) the modal value of the standard (or typical) claims. 
However, it is observed that a small number of outlying passenger claims are approximately a multiple $\beta$ times higher than those whose claims 
are near the modal value. 
A random sample of size 25 of the claim amounts from the year 2009 is available, and is given below: 

\vspace{0.25in} 
\indent
\ \ \ \ \ \ \ \ \ 280,870, \ \ \ 110,147, \ \ \ 100,483, \ \ \ 108,729, \ \ \ 142,800 \\
\indent
\ \ \ \ \ \ \ \ \ 102,108, \ \ \ 107,852, \ \ \ 163,073, \ \ \ 118,722, \ \ \ 108,948, \\
\indent
\ \ \ \ \ \ \ \ \ 117,307, \ \ \ 180,237, \ \ \ 115,422, \ \ \ 123,086, \ \ \ 113,936, \\
\indent
\ \ \ \ \ \ \ \ \ 221,617, \ \ \ 112,211, \ \ \ 106,790, \ \ \ 178,104, \ \ \ 101,561, \\
\indent
\ \ \ \ \ \ \ \ \ 104,325, \ \ \ 110,343, \ \ \ 112,843, \ \ \ 131,537, \ \ \ 138,744. \\

\noindent

For this illustrative Bayesian analysis we adopt the same prior specification as in the example in Section 4. 
That is, the parameter $\alpha$ is assigned a diffuse $\text{gamma}( 0.001, 0.001 )$ prior distribution, $\beta$ is assigned a shifted 
exponential prior distribution as in \eqref{betaprior1}, and $\epsilon$ is assigned a beta$( 0.1842, 3.5 )$ prior distribution. 
As $\theta$ is unknown, it also requires the assignation of a prior distribution. 
We assume that local experts can say something informative about the amount of a minimum reasonable claim, and suppose for the purpose of this 
example that this is well described by assigning $\theta$ a $\text{gamma}( 10, 0.0001 )$ prior distribution. 
The mean of this distribution is 100,000 and its standard deviation is approximately 31,623. 
It assigns prior probability of approximately 90\% to the interval between 50,000 and 150,000. 

\begin{figure}[!th]
\centering
\includegraphics[width=1.0\textwidth]{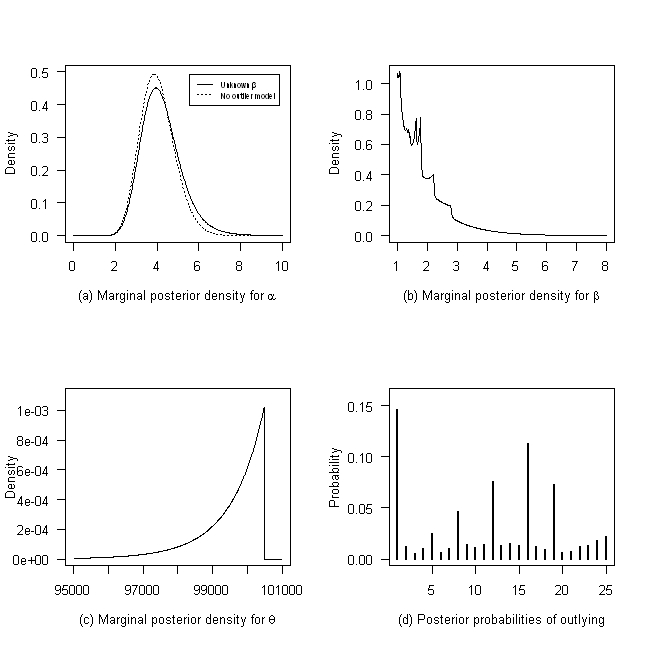}
\caption{Marginal posteriors for $\alpha$, $\beta$, $\theta$, and $\boldsymbol{\delta}$.}
\label{medical1} 
\end{figure}

\begin{figure}[!th]
\centering
\includegraphics[width=1.0\textwidth]{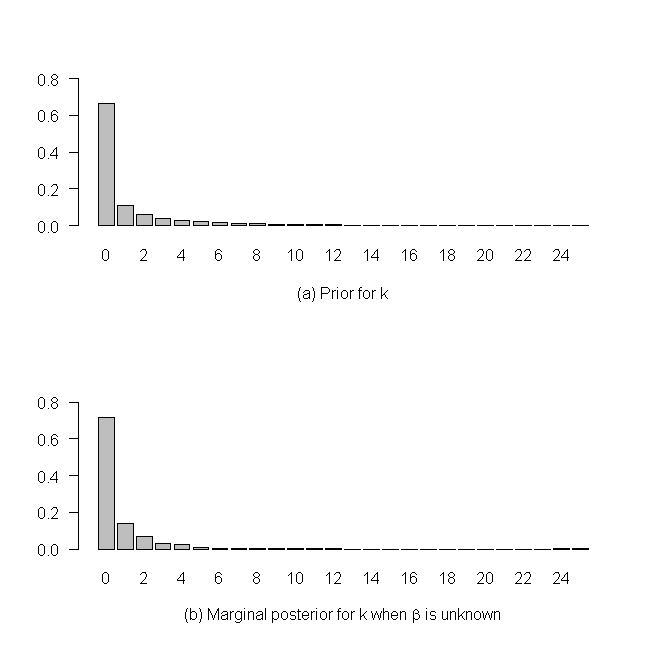}
\caption{Marginal prior and posterior probability mass functions for k.}
\label{medical1k}
\end{figure}

Results of the Gibbs sampler based Bayesian analysis of the model and data are displayed in Figures \ref{medical1} and \ref{medical1k}. 
Figure \ref{medical1}(a) also displays the posterior marginal density for $\alpha$ that results when the basic Pareto model \eqref{model1} 
(i.e. a Pareto model with no special accommodation made for outliers) is applied to the data in conjunction with the same prior for $\alpha$ 
and $\theta$ as before. 
As in the previous examples, it is clear that the Pareto scale-inflated outlier model shifts more of the posterior mass towards larger values 
of $\alpha$ so as to be more in line with the underlying distribution or value of $\alpha$ driving the amount of the standard claims. 
This should result in more sensible estimates and more reasonable and accurate statements about the predictive distribution of future standard 
medical claim amounts. 
Table 4 lists a number of the quantile values for the predictive claim amount distributions associated with the standard vehicles that result 
under the scale-inflated outlier and basic Pareto models.

\section{Funding}

This work was supported by a grant from the Natural Sciences and Engineering Research Council of Canada (NSERC).

\footnotesize
\begin{table}[tbp]
\begin{center}
\centerline{Table 4.} 
\centerline{Quantiles of the predictive distribution for standard medical claims under various models.} 
\vspace*{.1in}
\begin{tabular}{ @{\extracolsep{\fill}} | c c c c c | @{\extracolsep{0pt}} } 
\hline 
\, Pareto model \, & Median & 75\% & 90\% & 95\% \\ 
\hline
Scale-shifted & 882,127 & 1,571,983 & 3,408,270 & \, 6,267,162 \, \\ 
Basic & 912,938 & 1,680,812 & 3,757,930 & 6,909,201 \\ 
\hline 
\end{tabular} 
\end{center}
\end{table} 
\normalsize

\def \baselinestretch{1.5}

\newpage

\vskip.8in
\def \baselinestretch{1.2}
\large \normalsize

\begin{center}
\section*{\large REFERENCES} 
\end{center}

\def\beginref{\begingroup
              \clubpenalty=10000
              \widowpenalty=10000
              \normalbaselines\parindent 0pt
              \parskip.5\baselineskip
              \everypar{\hangindent3em}}
\def\endref{\par\endgroup}
\beginref

Arnold, B.C., \textit{Pareto distributions}, Fairland, MD: International Co-operative Publishing House, 1983. 


de Alba, E., ``Claims Reserving When There Are Negative Values in the Runoff Triangle: Bayesian Analysis Using the Three-Parameter 
Log-Normal Distribution," \textit{North American Actuarial Journal} 10(3), 2006, 45-59.

Dixit, U.J., Jabbari Nooghabi, M., ``Efficient estimation in the Pareto distribution with the presence of outliers," 
\textit{Statistical Methodology} 8, 2011a, 340-355. 

Dixit, U.J, Jabbari Nooghabi, M., ``Efficient Estimation of the Parameters of the Pareto Distribution in the Presence of Outliers," 
\textit{Communications of the Korean Statistical Society} 18, 2011b, 817-835. 

Gelman, A., Carlin, J.B., Stern, H.S., and Rubin, D.B., \textit{Bayesian Data Analysis}, 2nd ed., 
New York: Chapman \& Hall/CRC, 2004. 

Grubbs, F.E., ``Procedures for detecting outlying observations in samples," \textit{Technometrics} 11, 1969, 1-21. 

Guttman, I., Dutter, R. and Freeman, P., ``Care and handling of univariate outliers in the general linear model to detect
spuriosity - a Bayesian approach," \textit{Technometrics} 20, 1978, 187-194.

Guttman, I., ``Care and handling of univariate or multivariate outliers in detecting spuriosity - a Bayesian approach," 
\textit{Technometrics} 5, 1973, 723-738.

Johnson, N.L., Kotz, S., and Balakrishnan, N., \textit{Continuous univariate distributions}, Vol. 1, 2nd ed., 
New York: John Wiley, 1994. 

McNeil, A.J., ``Estimating the tails of loss severity distributions using extreme value theory," \textit{ASTIN Bulletin} 27, 1997, 
117-137. 

Ntzoufras, I., \textit{Bayesian Modeling Using WinBUGS}, Hoboken, NJ: John Wiley \& Sons, Inc., 2009. 

Ntzoufras, I. , Katsis, A. and Karlis, D.,``Bayesian Assessment of the Distribution of Insurance Claim Counts Using Reversible 
Jump MCMC," \textit{North American Actuarial Journal} 9, 2005, 90-108. 

Rosenbaum, F., \textit{Mathematical Concepts in the Insurance Industry}, Swiss Reinsurance Company, 2011. 

Schmutz, M., and Doerr, R.R., \textit{The Pareto model in property reinsurance - Formulas and applications with reference to other loss 
distribution functions}, Swiss Reinsurance Company, Zurich, 1998. 

Scollnik, D.P.M., ``On composite Lognormal-Pareto models", \textit{Scandinavian Actuarial Journal}, 2007, 20-33.

Scollnik, D. P. M., ``Comments on two papers concerning estimation of the parameters of the Pareto distribution in the presence 
of outliers", \textit{Statistical Methodology}, In Press. 

Verdinelli, I., and Wasserman, L., ``Bayesian analysis of outlier problems using the Gibbs sampler," 
\textit{Statistics and Computing}, 1991, 105-117.

Verrall, R. J., ``A Bayesian Generalised Linear Model for the Bornhuetter-Ferguson Method of Claims Reserving," 
\textit{North American Actuarial Journal} 8(3), 2004, pp. 67-89.

Verrall, R.J., ``Obtaining Predictive Distributions for Reserves Which Incorporate Expert Opinion," \textit{Variance} 1(1), 2007, 53-80

\newpage 

\newpage 

\newpage 

\newpage 

\newpage 

\newpage 

\newpage 

\newpage 

\end{document}